\documentclass[prd,twocolumn,tightenlines,superscriptaddress,nofootinbib,
showpacs]{revtex4}
\usepackage{amssymb,latexsym}
\usepackage{amsmath,amsbsy,bbm}
\usepackage{epsfig,bm}
\usepackage{graphicx,comment}
\usepackage{color}
\usepackage{soul}
\usepackage[normalem]{ulem}
\begin{document}

\title{Universal quantum criticality at finite temperature
for two-dimensional disordered and clean dimerized spin-$\frac{1}{2}$ antiferromagnets }

\author{D.-R. Tan}
\affiliation{Department of Physics, National Taiwan Normal University,
88, Sec.4, Ting-Chou Rd., Taipei 116, Taiwan}
\author{F.-J. Jiang}
\email[]{fjjiang@ntnu.edu.tw}
\affiliation{Department of Physics, National Taiwan Normal University,
88, Sec.4, Ting-Chou Rd., Taipei 116, Taiwan}

\begin{abstract}
The quantum critical regime (QCR) of a two-dimensional (2D) disordered and 
a 2D clean dimerized spin-$\frac{1}{2}$ Heisenberg models are studied
using the first principles nonperturbative quantum Monte Carlo simulations 
(QMC). In particular, the three well-known universal 
coefficients associated with QCR are investigated in detail. 
While in our investigation we find the obtained results are consistent with the 
related analytic
predictions, non-negligible finite temperature ($T$) effects are observed. 
Moreover, a striking finding in our study 
is that the numerical value for one of the universal coefficients we determine 
is likely to be different significantly from the corresponding (theoretical)
result(s) established in the 
literature. To better understand the sources for the discrepancy observed here, 
apart from carrying out the associated analytic calculations not considered
previously, it will be desirable as well to conduct a comprehensive examination 
of the exotic features of QCR for other disordered and clean spin systems than 
those investigated in this study.

\end{abstract}


\maketitle

\section{Introduction}

The two-dimensional (2D) quantum antiferromagnets, both with and without 
charge carriers,
are among the most important systems in condensed matter physics. 
From the experimental perspective, these materials are related to
the high temperature (high $T_c$) cuprate superconductors. As a result,
numerous associated experiments were conducted and the obtained data
have triggered many theoretical studies of these systems, including accurate 
determination of their low-temperature properties such as the 
staggered magnetization density and the spin stiffness 
\cite{Cha88,Reg88,Cha89,Oit941,Oit942,Wie94,San951,Bea96,San97,Jia08,Jia11.1}.    

Theoretically, at zero temperature and in the ordered phase, 
the 2D spin-$\frac{1}{2}$ Heisenberg
antiferromagnet can be treated classically and this region is 
known as the renormalized classical regime in the literature. 
When the long-range order of the system is destroyed 
by the quantum fluctuations, a completely different portrait
of its ground states called the quantum disordered regime appears. 
Moreover, in both regimes, as the temperature $T$
rises, there will be crossovers such that the system enters yet another
unique phase called the quantum critical regime (QCR). 
In particular, due to the interplay between
the thermal and the quantum fluctuations, some exotic
characteristics will emerge in QCR. These special features of 
QCR is signaled out by the presence of several universal 
behavior among some physical quantities of the underlying 2D spin 
system \cite{Chu93,Chu931,Sok94,Chu94}.

Based on the analytic calculations using the method of large-$N$ expansion 
for the effective nonlinear sigma model of the 2D Heisenberg antiferromagnet, 
three universal relations are established (assuming the dynamic 
exponent $z$ is 1): $\chi_u = \frac{\Omega}{c^2} T$, 
$S(\pi,\pi)/\chi_s = \Xi T$, and $c/\xi = X T $. Here $\chi_u$, $c$, 
$S(\pi,\pi)$, $\chi_s$, and $\xi$
are the uniform susceptibility, the 
spinwave velocity, the staggered structure factor, the staggered susceptibility, 
and the correlation length, respectively. Moreover, the coefficients
$\Omega$, $\Xi$, and $X$ are universal, namely their numerical values are 
independent of any microscopic details. For 2D dimerized Heisenberg 
models with spatial anisotropy, QCR as well as the related universal 
coefficients should be detectable at any values of the corresponding
tuning parameter. It is probable as well that systems with (certain kinds of) 
quenched disorder may exhibit features of QCR. 

Interestingly, while universal behavior characterizing QCR 
is indeed observed for several numerical studies of 2D dimerized spin models,
crystally clear evidence only found at the finite temperature regions above 
the related quantum critical points (QCPs). In other words, when the
associated calculations are conducted away from QCPs, 
the emergence of the exotic QCR scaling has not been established 
rigorously and numerically yet. 
\cite{San95,Tro96,Tro97,Tro98,Kim99,Kim00}.
For example, as introduced in the previous paragraph regarding QCR, 
a plateau is supported to appear in a certain region of the inverse temperature $\beta$ 
when $S(\pi,\pi)/(\chi_s T)$ is treated as a function of $\beta$. However, such
a scenario does not occur in the relevant studies when the used data were determined
away from the corresponding QCPs. 
 
At the moment, numerical studies related to QCR have been focusing on 
clean dimerized spin systems. The exploration of whether features of QCR, 
in particular the three universal quantities mentioned previously, 
exist for models with quenched disorder have been examined only implicitly, not 
systematically. Motivated by this, here we simulate a 2D
spin-$\frac{1}{2}$ Heisenberg model on the square lattice with a kind of
(quenched) disorder using the quantum Monte Carlo (QMC) calculations. 
The employed disorder distribution is based on the so-called
configurational disorder introduced in Ref.~\cite{Yao10}.  
Apart from this, the 2D clean dimerized plaquette 
quantum spin system is investigated as well for comparison and
clarification.

Remarkably, for both the considered disordered and clean models, features
of QCR do emerge at their corresponding QCPs. Furthermore, 
non-negligible $T$ dependence for the universal quantities of QCR, which was 
overlooked before, is found in our investigation. The most
striking result obtained here is that, the numerical values of
the universal coefficient $\Omega$ determined in our study for both the 
considered models are likely to deviate significantly from those 
calculated previously in Refs.~\cite{Chu94,Tro97,Tro98,Kim00}.
The evidence provided here for the described 
variation regarding the numerical value of $\Omega$ is convincing. 
This finding of ours is consistent with that obtained in Ref.~\cite{Sen15} for which the 
conclusion is based on a detailed study of a 2D clean bilayer quantum spin system.
To better understand the sources of the discrepancy found here, 
apart from carrying out the analytic calculations associated with 
corrections not taken into account previously, a more thorough exploration of other disordered 
and clean spin models than those studied here is desirable. 

The rest of the paper is organized as follows. After the introduction,
the considered models as well as the required physical quantities for
investigating the features of QCR are described.
A detailed analysis, focusing on the three well-known universal
coefficients of QCR, is presented then. In particular, the numerical evidence
for the discrepancy mentioned above is given. Finally, a section is devoted to 
conclude our study shown here.

\section{Microscopic model and observables}
\begin{figure}
\begin{center}
\vbox{
\includegraphics[width=0.3\textwidth]{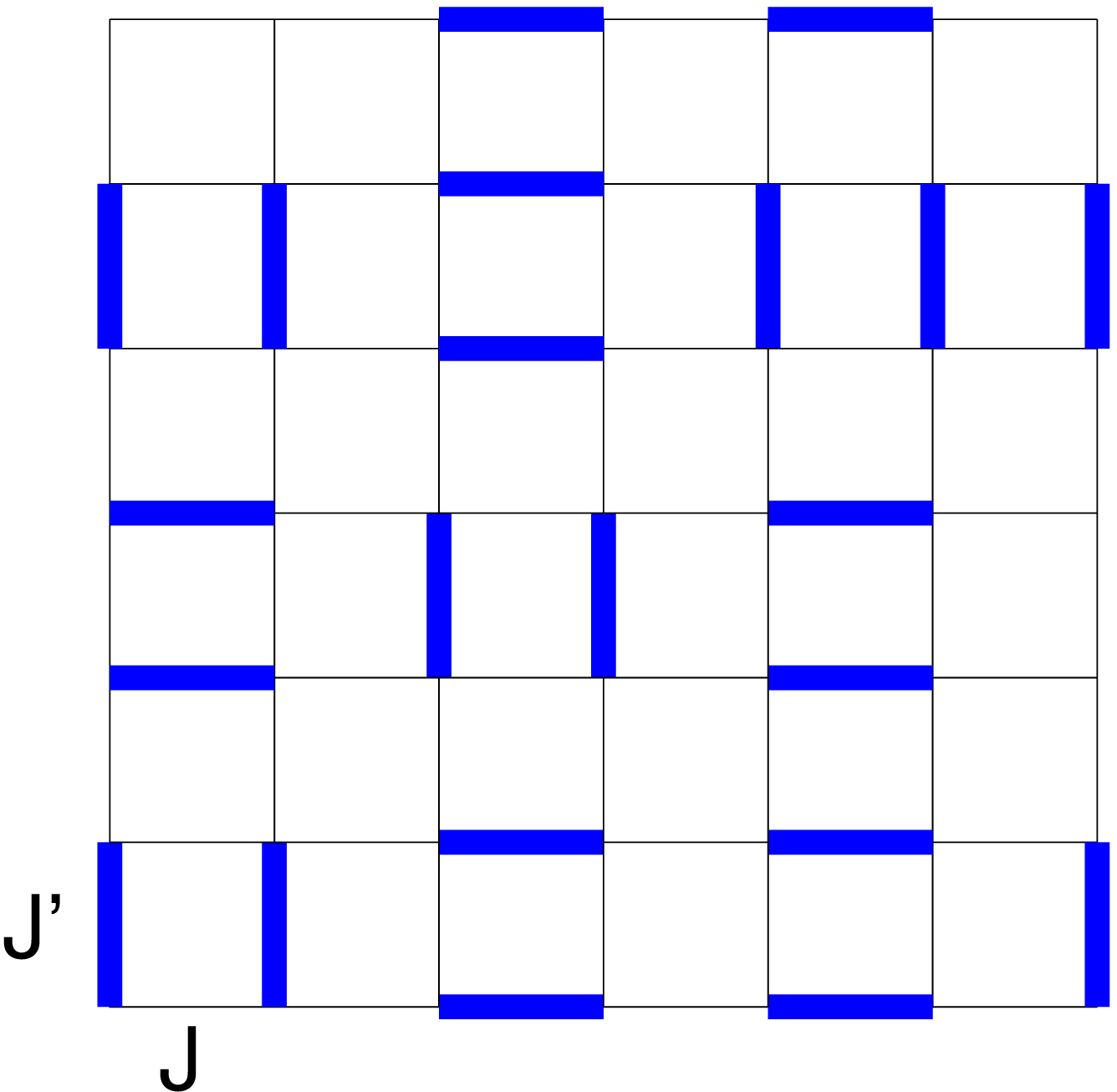}
\vskip0.7cm
\includegraphics[width=0.27\textwidth]{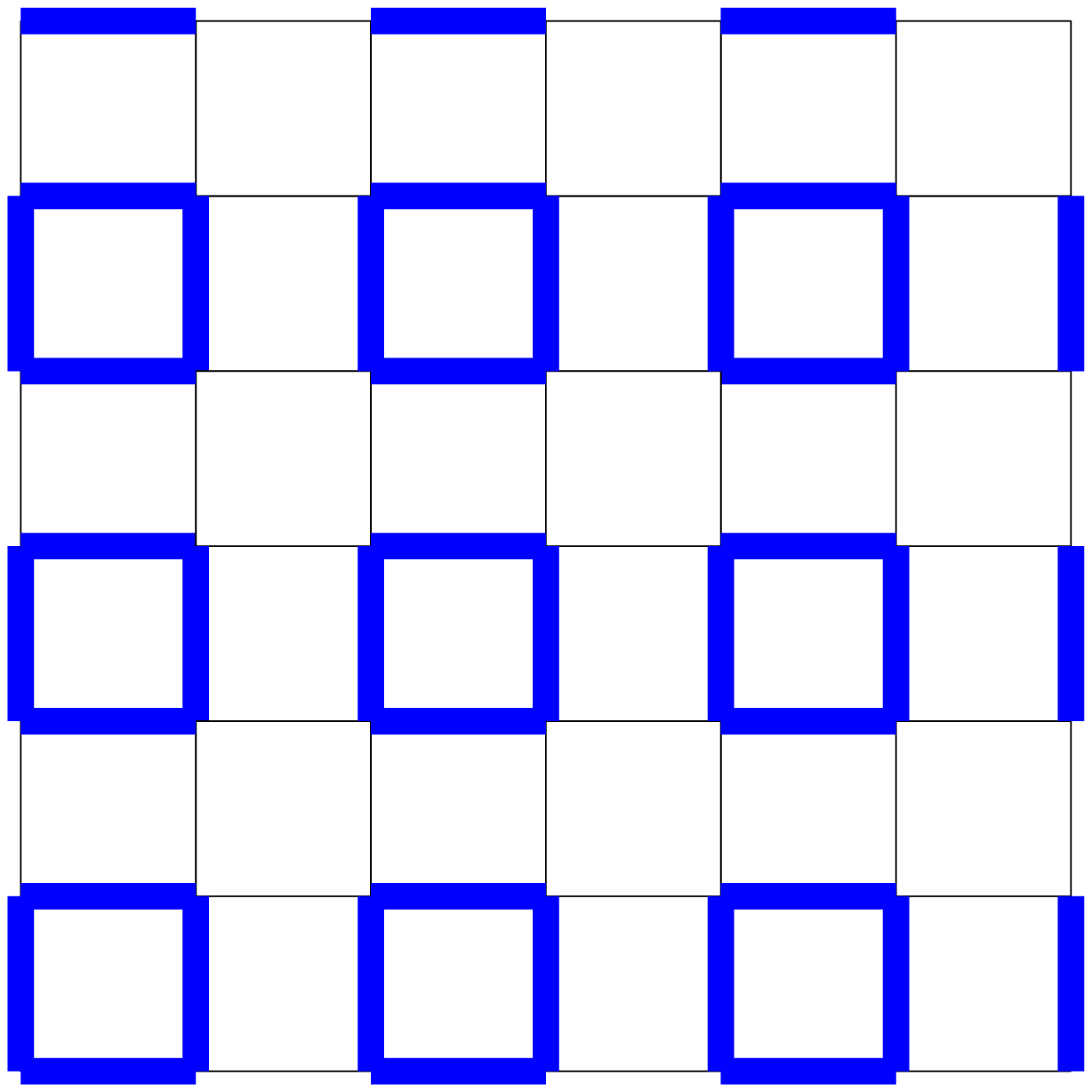}
}\vskip0.25cm
\end{center}
\caption{The model with configurational disorder (top panel)
and the clean dimerized plaquette model (bottom panel) considered in this study.}
\label{model_fig1}
\end{figure}

The 2D spin-$\frac{1}{2}$ system with a quenched disorder and the 2D 
clean quantum dimerized plaquette Heisenberg model 
studied here are given by the same form of Hamilton operator
\begin{eqnarray}
\label{hamilton}
H &=& \sum_{\langle ij \rangle}J_{ij}\,\vec S_i \cdot \vec S_{j} 
+ \sum_{\langle i'j' \rangle}J'_{i'j'}\,\vec S_i \cdot \vec S_{j}, \\
\end{eqnarray}
where $J_{ij}$ and $J'_{i'j'}$ are the antiferromagnetic 
couplings (bonds) connecting nearest neighbor spins $\langle  ij \rangle$ 
and $\langle  i'j' \rangle$, respectively, 
and $\vec{S}_i$ is the spin-$\frac{1}{2}$ operator at site $i$.
The quenched disorder considered in this investigation
is based on the idea of configurational disorder \cite{Yao10}. Specifically,  
in our simulations for the disordered model, the probabilities of 
putting a pair of $J'$-bonds vertically and horizontally in a plaquette 
consisting of two by two spins are both 0.5. (We will still use the term
configurational disorder for this employed disorder distribution 
in the rest of the paper).
Figure~\ref{model_fig1} demonstrates the studied disordered and clean models. 
Here we use the convention that the couplings $J'$ and 
$J$ satisfy $J' > J$. As a result, each of the considered system will undergo 
a quantum phase transition once the ratio $J'/J$ exceeds a particular value
called the critical point. These special points in the associated parameter 
spaces are commonly denoted by $(J'/J)_c$ in the literature.

To examine the well-known universal relations of QCR, 
particularly to understand whether these relations appear for the considered
disordered system, the staggered structure factor $S(\pi,\pi,L)$, the
uniform and staggered susceptibilities ($\chi_u$ and $\chi_s$), the
spinwave velocity $c$, as well as the correlation length $\xi$ are measured.
The staggered structure factor $S(\pi,\pi,L)$ on a finite lattice with a
linear box size $L$ is defined by
\begin{equation}
S(\pi,\pi,L) = 3 L^2 \left\langle ( m_s^z )^2\right\rangle,
\end{equation}
where $ m_s^z = \frac{1}{L^2}\sum_{i}(-1)^{i_i + i_2}S_i^z$ and the summation is over all sites.
The uniform susceptibility $\chi_u$ and staggered susceptibility $\chi_s$ take the forms
\begin{equation}
\chi_u = \frac{\beta}{L^2}\left\langle\left(\sum_{i}S_i\right)^2 \right\rangle
\end{equation}
and 
\begin{equation}
\chi_s = 3L^2\int_0^{\beta} \langle m_s^z(\tau) m_s^z(0)\rangle d\tau,
\end{equation}
respectively. The quantity $\beta$ appearing above is the inverse temperature. 
In addition, the correlation length $\xi$ is expressed as
\begin{eqnarray}
\xi &=& \frac{L_1}{4\pi}\sqrt{\frac{S(\pi,\pi)}{S(\pi+2\pi/L_1,\pi)}-1} \nonumber \\
&& + \frac{L_2}{4\pi}\sqrt{\frac{S(\pi,\pi)}{S(\pi,\pi+2\pi/L_2)}-1},
\end{eqnarray}
where the quantities $S(\pi+2\pi/L_1,\pi)$ and $S(\pi,\pi+2\pi/L_2)$ are the 
Fourier modes with the second largest magnitude.
Finally the spinwave velocities $c$ for both the investigated models are 
calculated through the temporal and spatial winding numbers squared 
($\langle W_t^2\rangle$ and $\langle W_i^2 \rangle$ with $i\in\{1,2\}$).

We would also like to point out that while the same notations are used
here for both the observables of the two studied spin systems, whenever a
physical quantity associated with the disordered model is
presented, it is obtained by averaging over the generated configurations
of randomness.

\section{The numerical results}

To study the features of QCR associated with the considered disordered
and clean models, we have performed large scale QMC simulations
using the SSE algorithm with very efficient operator-loop update 
\cite{San99}. For the disordered quantum spin system, while most of
the corresponding results presented here are obtained by averaging over 
several hundred realizations of randomness, the outcomes related to
the spinwave velocity are calculated using (a) few thousand disorder 
configurations. 

For a given $J'/J$ and for the corresponding results of finite $T$, a generated configuration 
of randomness is employed for the calculations associated with several
sequential values of $\beta$. Furthermore, for (almost) every considered temperature 
at least five thousand Monte Carlo (MC) sweeps as well as the step of adjusting 
cut-off in the SSE algorithm are carried out for both the processes of 
thermalization and measurement. Therefore the correlations among the obtained 
data are anticipated to be mild. In particular, the first few data in a Monte 
Carlo simulation are disregarded for the disorder average. As a result, the 
potential issue of thermalization in studies of disordered systems is under 
control. Indeed, the outcomes resulting from several additional calculations 
using 2500 MC sweeps for the thermalization agree remarkably well with those 
explicitly shown here.
  
We would also like to emphasize the fact that the uncertainties of the 
calculated observables should be dominated by the number of configurations 
used for the disorder average. This is because for each considered set 
of parameter, the number of MC sweeps employed for the related simulations is 
much larger than that of the associated configurations generated. Still, the 
errors for the obtained quantities are estimated with conservation so that 
the statistical uncertainties resulting from the MC simulations are not 
overlooked. In addition, most (a few) of the results presented here are obtained
on $L=256$ ($L > 256$) lattices.
For comparison, some outcomes determined with smaller box sizes are shown
as well.

\begin{figure}
\vskip0.75cm
\begin{center}
\includegraphics[width=0.35\textwidth]{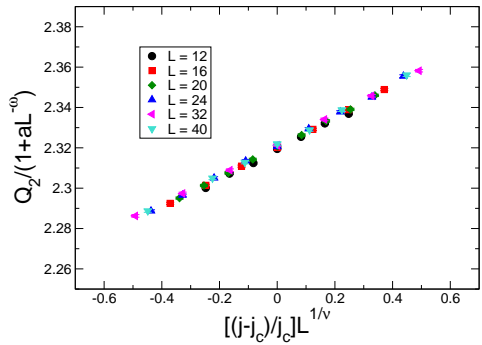}
\vskip0.25cm
\end{center}
\caption{$Q_2/(1+aL^{-\omega})$ (of the studied disordered model) as functions of 
$[(j_c-j)/j_c]L^{1/\nu}$ for various $L$ and $J'/J$.
Here $j$ and $j_c$ are defined as $j = J'/J$ and $j_c = (J'/J)_c$, 
respectively. While the numerical values of the 
coefficients $a = -0.5$, $\omega = 1.0$, and $j_c = 1.990$ are taken directly
from the outcomes
determined in Ref.~\cite{Yao10}, the $\nu$ used in producing the universal 
curve is its theoretical prediction $\nu = 0.7115$.}
\label{Jc_fig1}
\end{figure}

To carry out a comprehesive (and detailed as well) study of QCR for the
investigated models, ideally the calculations should be conducted
at the associated QCPs.  
In theory, close to a second order quantum phase transition and for various 
$L$ and $(J'/J)$, if one treats the results of data collapse of 
$Q_2/(1+aL^{-\omega})$ as functions of $\left[(j-j_c)/j_c\right]L^{1/\nu}$, 
then a universal smooth curve should emerge. 
Here $j = J'/J$, $j_c = (J'/J)_c$, $\nu$ and $\omega$ are 
the correlation length and the confluent critical exponents, respectively, 
and $a$ is some constant. Moreover, $Q_2$ is the second Binder ratio
which is defined by 
$Q_2 = \frac{\langle (m_s^z)^4 \rangle}{\langle (m_s^z)^2\rangle^2}$.
Interestingly, the zero temperature data we calculate with the $\beta$-doubling
scheme \cite{San02} are 
fully consistent with the outcomes reached in Ref.~\cite{Yao10}.
Indeed, using $a=-0.5$, $\omega = 1.0$, $j_c = 1.990$ (these three
results are taken directly from Ref.~\cite{Yao10}), $\nu = 0.7115$ (the established
value for this exponent), as well as the data obtained here,
we have reproduced the associated universal curve of $Q_2$ just 
like the one shown in Ref.~\cite{Yao10}, see Fig.~\ref{Jc_fig1}.
With a fixed $\nu=0.7115$, we additionally fit the observables
$Q_2$ and $\rho_s L$ \cite{rhosL_def} to their expected scaling formulas near 
QCP. The $(J'/J)_c$ determined from these fits associated with $Q_2$
and $\rho_s L$ are both given by $(J'/J)_c = 1.990(1)$. This strongly suggests 
that taking $1.990(1)$ as the QCP for the investigated disordered model
should be beyond any doubt. 

The fact that the $(J'/J)_c$ resulting from 
$Q_2$ and $\rho_s L$ agree quantitatively with each other indicates
that the dynamic exponent $z$ related to the considered disordered system is
1. We will demonstrate shortly that this is truly the case. 

Finally, the QCP of the clean
plaquette model has been calculated in Ref.~\cite{Wen09} and is given by 
$(J'/J)_c = 1.8230(3)$.

\subsection{The determination of spinwave velocity $c$}
The spinwave velocities $c$ at the critical points for the studied models
are calculated
using the method of winding numbers squared proposed in Refs.~\cite{Jia11,Sen15}.
Specifically, for a fixed box size $L$ (and a fixed $J'/J)$, the value of 
$\beta$ is adjusted in the calculations so that the temporal winding number 
squared $\langle W_t^2\rangle$ agrees quantitatively with that of the 
averaged spatial winding numbers squared 
$\langle W^2\rangle = \frac{1}{2}\sum_{i=1,2}\langle W_i^2\rangle$. 
Under such a condition, the corresponding spinwave velocity $c(L,J'/J)$ is
determined by the equation $c(L,J'/J) = L/\beta^{\star}$, where 
$\beta^{\star}$ is the inverse temperature for which the condition described
above regarding the winding numbers squared is fulfilled.  
Since this method is valid only when the long-range
antiferromagnetic order is present in the system, the relevant
simulations are done at $J'/J = 1.988$ for the disordered system
(which has $(J'/J)_c = 1.990(1)$ \cite{Yao10}). 
For the clean plaquette model, calculations at several selected
$J'/J < (J'/J)_c = 1.8230(3)$ , as well as fits and interpolations 
are conducted in order to obtained the bulk $c$ at the
associated critical point. 

\subsubsection{The spinwave velocity of the disordered system}
\begin{figure}
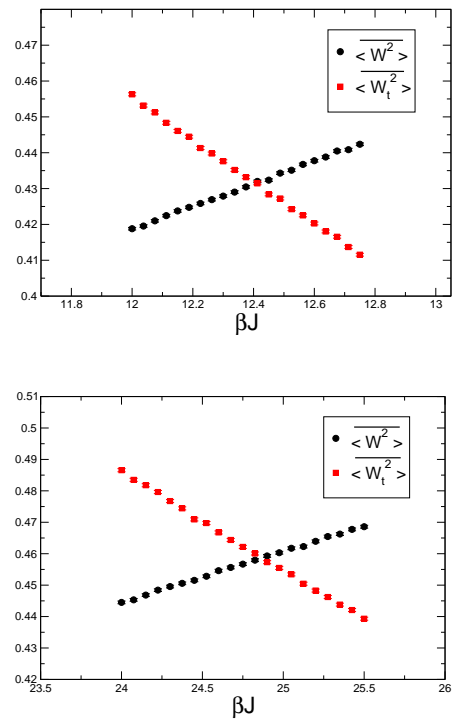

\begin{center}
\vbox{
\includegraphics[width=0.325\textwidth]{config_c_L24.eps}
\vskip0.7cm
\includegraphics[width=0.325\textwidth]{config_c_L48.eps}
}\vskip0.25cm
\end{center}
\caption{The temporal and spatial winding numbers squared as functions of 
$\beta$ for the studied disordered model. The simulations are conducted
at $J'/J = 1.988$ and the results shown in the 
top and the bottom panels are determined with $L=24$ and 
$L=48$, respectively.}
\label{c_fig1}
\end{figure}

The $\langle W_t^2\rangle$ and $\langle W^2 \rangle$ as functions of $\beta$
for the studied disordered system are shown in Fig.~\ref{c_fig1}. 
The calculations are done at $J'/J = 1.988$ and the outcomes
presented in the top and bottom panels of the figure are obtained 
with $L=24$ and $L=48$, respectively. The corresponding
values of $c$ estimated conservatively from these two simulated results match
each other nicely. Indeed, while the calculated result of $c$ for $L = 24$
is given by $c = 1.934(5)J$, the $c$ determined from the data of $L=48$ is
found to be $c = 1.931(9)J$. We have additionally performed simulations
at $J'/J = 1.986$ with $L = 48$. The outcome of $c$ from the 
simulations associated with $J'/J = 1.986$ agrees remarkably well 
with that of $J'/J=1.988$. Therefore it should be accurate to use 
$c = 1.931(9)J$ as the bulk value of $c$ right at the critical point.

\subsubsection{The spinwave velocity of the clean plaquette model}
\begin{figure}
\vskip0.75cm
\begin{center}
\includegraphics[width=0.35\textwidth]{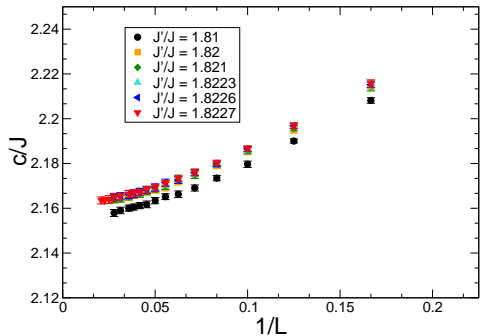}
\vskip0.25cm
\end{center}
\caption{The estimated $c$ as functions of $1/L$ for several selected 
$J'/J$ of the 2D plaquette model.}
\label{c_fig2}
\end{figure}

To determine the bulk $c$ at the critical point of the 2D clean plaquette 
model, a more thorough calculation is performed. In particular we carry out 
simulations with various box sizes $L$ at several selected $J'/J \le 1.8227$ 
close to the critical point $(J'/J)_c = 1.8230(3)$. The obtained results are 
shown in Fig.~\ref{c_fig2}. The bulk $c(J'/J)$ of each considered $J'/J$ is 
determined by applying the following two ans\"atze
\begin{eqnarray}
&&a_0 + a_1/L^2, \\
&&b_0 + b_1/L^2 + b_2/L^3
\end{eqnarray}
to fit the corresponding data. This strategy of calculating the bulk values
of $c$ is inspired by the results demonstrated in Ref.~\cite{Sen15}.
The uncertainty for the bulk $c$ of every used $J'/J$
is the standard deviation deriving from considering Gaussian noises 
in the associated weighted $\chi$-squared fits.   
With the outcomes from the fits employing ansatz (8), the $c$  
corresponding to $(J/'J)_c$ is estimated by interpolation based on a 
linear fit of the form $a(J'/J) + b$. 
With such a procedure, the spinwave velocity $c$ at the 
critical point is found to be $c = 2.163(4)J$. Here the quoted 
uncertainty is not determined directly from the interpolation, but is
calculated with conservation assuming that for $(J'/J)_c$ a similar 
statistic as those of the $J'/J$ shown in Fig.~\ref{c_fig2} 
is reached.

\subsection{The universal coefficient $\Omega$}
Theoretically the universal coefficient $\Omega$ is given by
$\chi_u = \left(\Omega/c^2\right)T^{d/z-1}$ at the critical point, 
where $d$ is dimensionality of
the system (which is 2 here) and $z$ is the associated 
dynamic exponent. The $z$ associated with the considered disordered model
is estimated to be 1 in Ref.~\cite{Yao10}. 

\subsubsection{The results of disordered system}
\begin{figure}
\begin{center}
\includegraphics[width=0.375\textwidth]{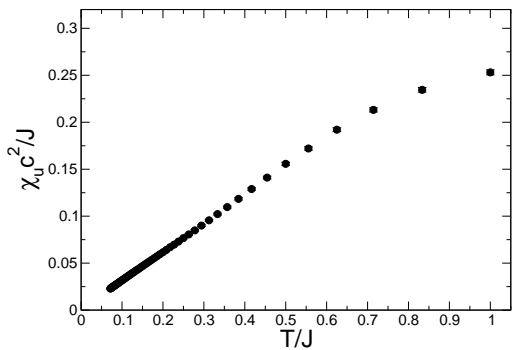}
\vskip0.25cm
\end{center}
\caption{$\chi_u c^2$ as a function of $T$ for the considered disordered model.
The data are obtained at the critical point $(J'/J)_c = 1.990$ with $L=256$.}
\label{omega_fig1}
\end{figure}

The $\chi_u c^2$ (determined at $(J'/J)_c = 1.990$ and on $L=256$ lattices) 
as a function of $T$ for the studied 2D spin-$\frac{1}{2}$ 
system with configurational disorder is depicted in Fig.~\ref{omega_fig1}.
Apart from the results obtained at the critical point $(J'/J)_c$,
we have additionally performed simulations at $J'/J = 1.989$ and 1.991
with $L=256$ so that for the considered observables the corresponding 
systematic errors due to the uncertainty of $(J'/J)_c$ can be investigated. 

The fits carried out here for the determination of $\Omega$ 
are done by fitting the data of $\chi_uc^2$ to both the ans$\ddot{a}$tze 
\begin{eqnarray}
&&a + b\,T^{2/z-1},\\ 
&&b_1\,T^{2/z_1-1}
\end{eqnarray}
with $a$, $b$, $b_1$, $z$, and $z_1$ left as the fitting parameters. With these two
formulas, the numerical values of $\Omega$ 
are just the parameters $b$ and $b_1$ calculated from the fits. In the 
following $z$ and $b$, instead of $z_1$ and $b_1$, will be used 
whenever the results from the fits employing the second ansatz are discussed, 
if no confusion arises.

\begin{figure}
\vskip0.3cm
\begin{center}
\vbox{
\includegraphics[width=0.375\textwidth]{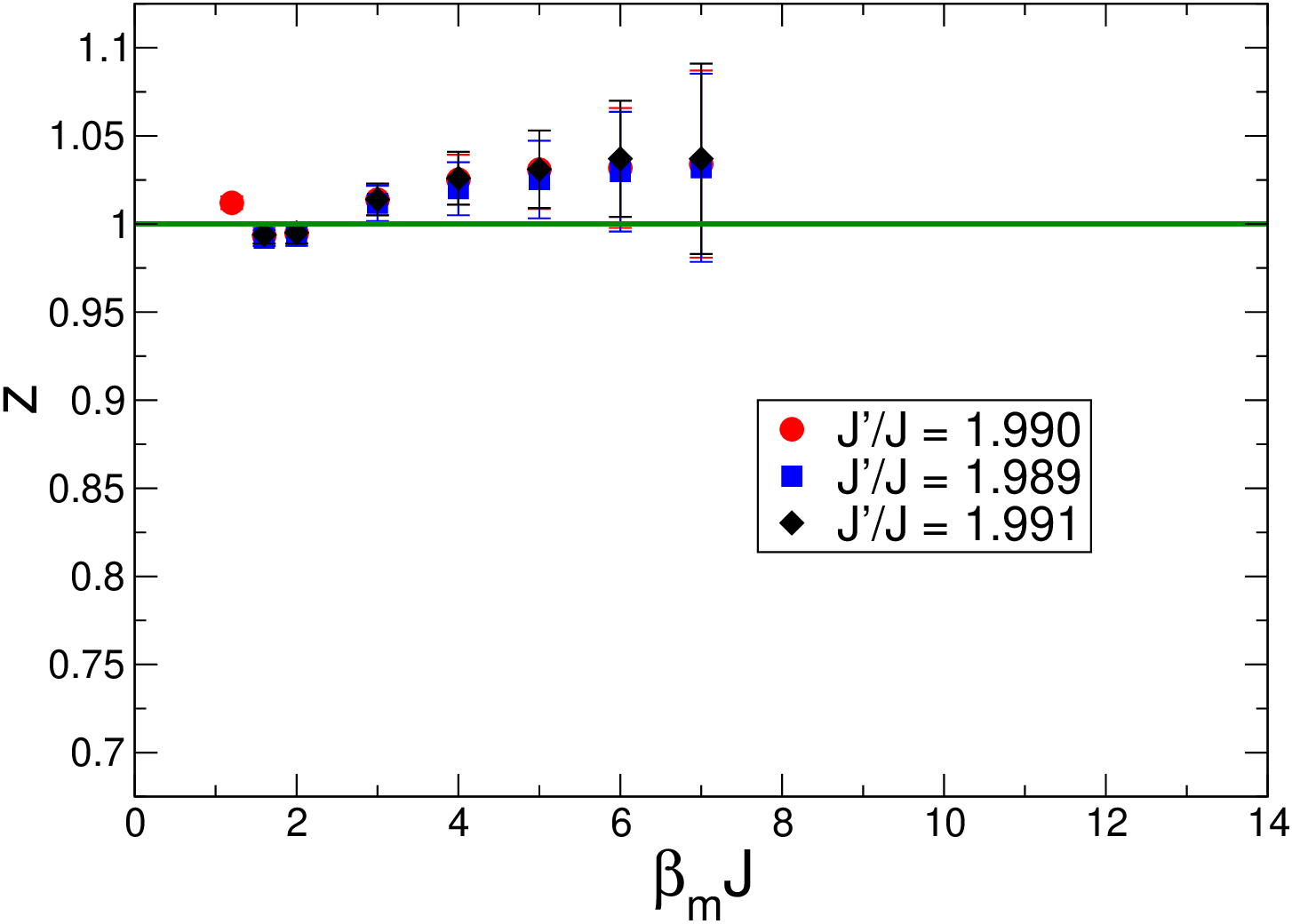}
\vskip0.85cm
\includegraphics[width=0.375\textwidth]{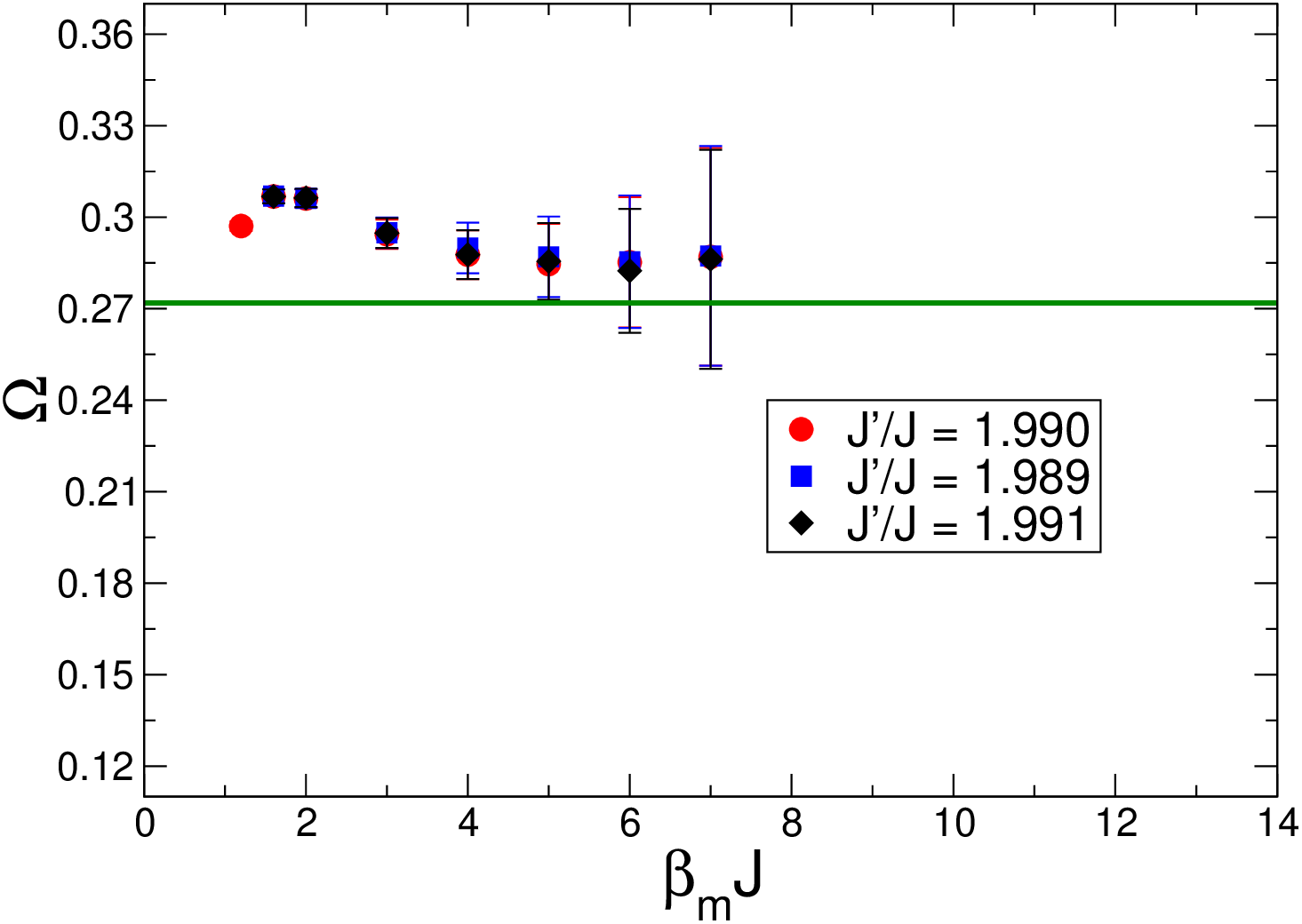}
}
\end{center}
\caption{The results of $z$ (top panel) )and $b$ ($\Omega$, bottom panel) for the considered disordered system. These
outcomes are obtained from the fits using the ansatz $a + b\,T^{2/z-1}$.
The horizontal ($x$) axes stand for the minimum values of $\beta$ used in the fits.
The solid lines in both panels are the corresponding theoretical predictions.}
\label{omega_fig2}
\end{figure}

The obtained results of $z$ and $b$ 
for all the three considered values of $J'/J$ are demonstrated in 
Figs.~\ref{omega_fig2} (using the first ansatz) and \ref{omega_fig3} 
(using the second ansatz). The horizontal ($x$) axes in these figures stand
for the minimum values of $\beta$ used in the fits.
Interestingly, as one can see from
the figures, most of the determined values of $z$ are slightly above 1. 
Moreover, the calculated $b$ are larger than 0.27185 (solid horizontal lines in 
the bottom panels of both Figs.~\ref{omega_fig2} and \ref{omega_fig3}). 
Although $b$ is approaching $0.27185$ when more data determined at high temperature
region are excluded in the associated fits using the first ansatz, the majority of 
the obtained results of $b$ are well above the corresponding theoretical 
prediction $\Omega = 0.27185$. Similar to the analysis for the spinwave velocity $c$,
here the errors shown in the figures are the standard deviations resulting from 
considering Gaussian noises in the related weighted $\chi$-squared fits. While not presented 
here, the $a$ determined from the fits are either with small magnitude (of the order 
$10^{-3}$) or are statistically identical to zero. 

It is intriguing to notice that when the first ansatz is considered, 
as the magnitude of the determined $z$ 
increases (This occurs when more and more data calculated at high temperatures 
are excluded in the fits), the value of $b$ obtained comes toward 0.27185. 
In other words, $z$ and $\Omega$ are correlated. Since the difference
between the $z$ found here and that estimated in Ref.~\cite{Yao10} 
is only at few percent level, it is unlikely that such deviations
are due to the fact that the $z$ calculated here for the studied disordered 
model is a
new one other than that found in \cite{Yao10}. 
Instead, the observed discrepancy should be treated as a result of 
not taking some corrections into account in the analysis. Indeed,
to the best of our knowledge, we are not aware of other formulas besides those
employed here for the fits. As we will demonstrate later,
such a scenario for $z$ and $\Omega$ occurs for the clean plaquette model
as well.

\begin{figure}
\vskip0.5cm
\begin{center}
\includegraphics[width=0.375\textwidth]{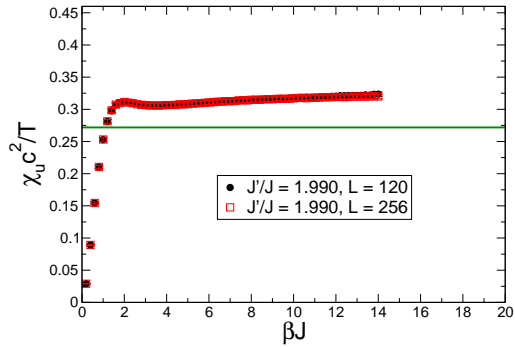}
\end{center}
\caption{$\chi_u c^2 / T$ as functions of $\beta$ for the studied
disordered model. 
The data are calculated at the critical point $(J'/J)_c=1.990$ with $L=120$
and $L=256$. The horizontal solid line is the theoretical prediction 
$\sim$ 0.27185.}
\label{omega_fig4}
\end{figure}

Since $z=1$ is beyond doubt for the considered disordered system, to accurately estimate 
the numerical value of $\Omega$, particularly to understand its dependence on 
(finite) temperature, it is helpful to investigate the quantity $\chi_u c^2/T$ as 
a function of the inverse temperature $\beta$. 
Such a study is inspired by the fact that a flat plateau should appear if the data of $\chi_u c^2/T$ 
are plotted against $\beta$ since $z=1$. Remarkably, a very flat plateau indeed emerges 
when $\chi_u c^2/T$ is treated as a function of $\beta$, 
see Fig.~\ref{omega_fig4}. While it is clear that the quantity 
$\chi_u c^2 / T $ receives mild corrections from terms taking some forms 
in $T$, the quality of flatness shown in Fig.~\ref{omega_fig4}
strongly indicates that the value of the universal coefficient $\Omega$ is 
larger than 0.27185 (which is the horizontal line in Fig.~\ref{omega_fig4}).
The $L=120$ data of $\chi_u c^2 /T$ obtained at $(J'/J)_c$
are demonstrated in Fig.~\ref{omega_fig4} as well. The quantitative agreement 
between the $\chi_u c^2 / T$ 
data of $L=120$ and $L=256$ rules out the possibility that the deviations 
of $\Omega$ and $z$ from their expected values are due to finite-size effects. 

\begin{figure}
\vskip0.5cm
\begin{center}
\includegraphics[width=0.375\textwidth]{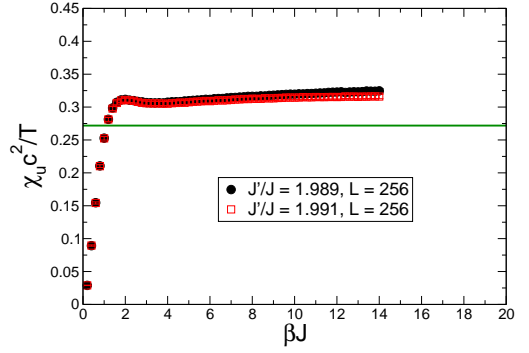}
\end{center}
\caption{$\chi_u c^2 / T$ as functions of $\beta$ for the studied
disordered model. 
The data are calculated at $J'/J = 1.991$ and 1.989 with $L=256$. The horizontal solid 
line is the theoretical prediction $\sim$ 0.27185.}
\label{omega_fig5}
\end{figure}

Aside from the results associated with $(J'/J)_c$, The $\chi_u c^2/T$ as functions of $\beta$ 
for both $J'/J = 1.989$ and 1.991 are shown in 
Fig.~\ref{omega_fig5}. As can been seen from the figure, flat plateaus well 
above $0.27185$ show up as well. The results presented in Fig.~\ref{omega_fig5} 
exclude the scenario that the observed discrepancy is due to the 
uncertainty of the critical point.

\begin{figure}
\vskip0.3cm
\begin{center}
\vbox{
\includegraphics[width=0.375\textwidth]{z_config.eps}
\vskip0.85cm
\includegraphics[width=0.375\textwidth]{omega_config.eps}
}
\end{center}
\caption{The results of $z$ (top panel) and $b$ ($\Omega$, bottom panel) for the considered disordered system. These
outcomes are obtained from the fits using the ansatz $b\,T^{2/z-1}$.
The horizontal ($x$) axes stand for the minimum values of $\beta$ used in the fits.
The solid lines in both panels are the corresponding theoretical predictions.}
\label{omega_fig3}
\end{figure}

\subsubsection{The results of clean system}

\begin{figure}
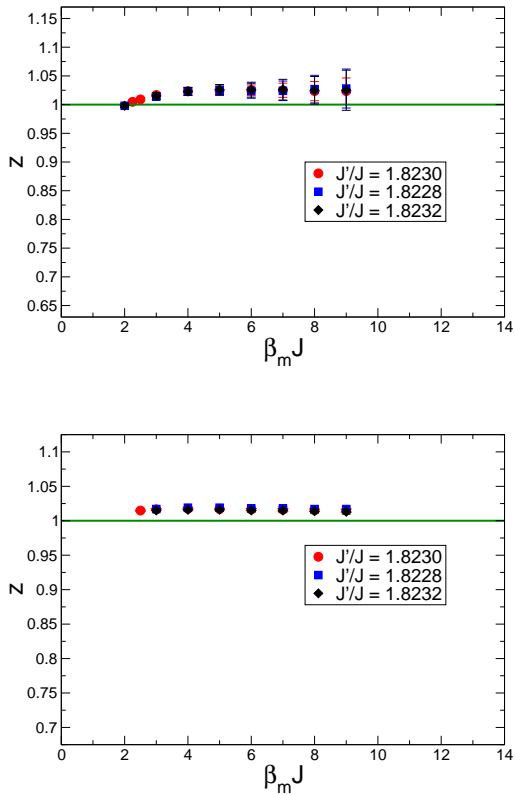

\vskip0.3cm
\begin{center}
\vbox{
\includegraphics[width=0.375\textwidth]{wc_z_plaq.eps}
\vskip0.85cm
\includegraphics[width=0.375\textwidth]{z_plaq.eps}
}
\end{center}
\caption{The results of $z$ for the clean plaquette model. These
outcomes are obtained from the fits using the ansatz $a + b\,T^{2/z-1}$ (top panel)
and $b_1\,T^{2/z_1-1}$ (bottom panel).
The horizontal ($x$) axes stand for the minimum values of $\beta$ used in the fits.}
\label{omega_fig6}
\end{figure}

While it is well-established that $z=1$ for the considered 2D plaquette model,
it will be useful to conduct a calculation like that done in the previous subsection 
to determine the dynamic exponent $z$ associated with the studied clean system.
Interestingly, a scenario like the one of the investigated disordered model
is observed. Specifically, the values of $z$ obtained here are slightly 
above the theoretical prediction $z=1$, see Fig.~\ref{omega_fig6}. 
Just like what has been argued previously, since the deviations
found are only at few percent level, these deviations should be treated as 
consequences resulting from (minor) corrections not taken into account in 
the analysis.

\begin{figure}
\vskip0.5cm
\begin{center}
\includegraphics[width=0.375\textwidth]{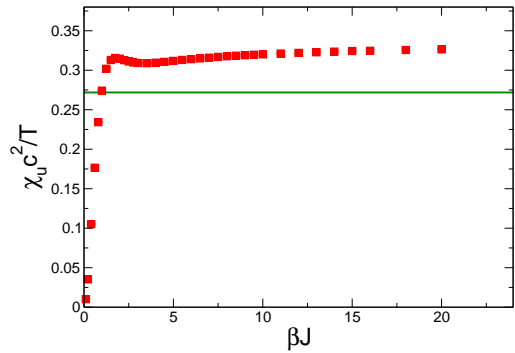}
\vskip0.25cm
\end{center}
\caption{Size-convergence $\chi_u c^2 / T$ as a function of $\beta$ for the studied
2D dimerized plaquette model.
The data are calculated at the critical point $(J'/J)_c=1.8230$ and the 
horizontal solid line is the theoretical prediction $\sim$ 0.27185.}
\label{omega_fig7}
\end{figure}

\begin{figure}
\vskip0.5cm
\begin{center}
\includegraphics[width=0.375\textwidth]{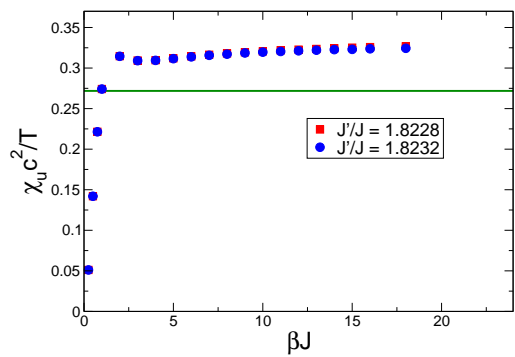}
\end{center}
\caption{Size-convergence $\chi_u c^2 / T$ as functions of $\beta$ for the studied
2D dimerized plaquette model.
The data are calculated at $J'/J=1.8228$ and 1.8232. The 
horizontal solid line is the theoretical prediction $\sim$ 0.27185.}
\label{omega_fig8}
\end{figure}

Similar to the analysis done in the previous subsection, we have also investigated 
the size-convergence quantity $\chi_u c^2 /T$ as a function of $\beta$ for the 
2D clean dimerized plaquette model \cite{size_convergence}. The considered data are determined at the 
expected critical point $(J'/J)_c = 1.8230$, as well as at $J'/J = 1.8228$, 1.8232
in order to take into account the effects from the uncertainties of $(J'/J)_c$.  
The resulting outcomes are depicted in Figs.~\ref{omega_fig7} and \ref{omega_fig8}.

Interestingly, while moderate $T$-dependence for $\chi_u c^2 /T$ definitely appears, 
as shown in the figures, one sees clearly that flat plateaus emerge as well.
Furthermore, by comparing the results presented in Figs.~\ref{omega_fig4}, \ref{omega_fig5}, 
\ref{omega_fig7} and \ref{omega_fig8}, the values of $\chi_u c/T^2$ for which all the plateaus 
take place match each other very well and are statistically above 0.27185.

In summary, the outcomes obtained here that $\Omega$ is quantitatively different
from its theoretical prediction 0.27185 is convincing.
In particular, based on the results of fits with a fixed $z = 1$ (in the formula (9)), the numerical value of $\Omega$ 
we estimate conservatively is about $0.306(10)$. A (slighly) larger number is reached if one uses the outcomes
calculated by considering only the lower temperature data for the fits. 
To arrive at a more accurate determination of $\Omega$ 
requires better understanding of its analytic expression. This is beyond the scope of our study
presented here.   

For the analysis done in the following (sub)sections, the assumption $z=1$ 
will be employed.

\vskip1.0cm

\begin{figure}
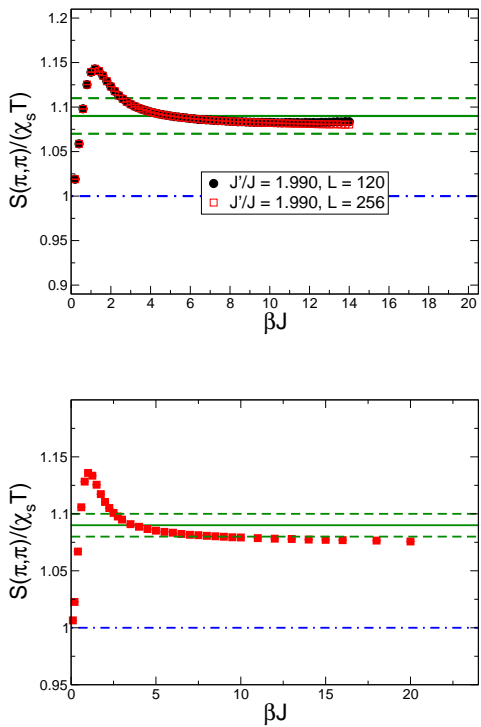

\begin{center}\vskip1.5cm
\vbox{
\includegraphics[width=0.35\textwidth]{J1.99_S_chis_T_beta.eps}
\vskip0.8cm
\includegraphics[width=0.35\textwidth]{plaq_S_chis_T_beta.eps}
}\vskip0.25cm
\end{center}
\caption{$S(\pi,\pi)/(\chi_s T)$ as functions of $\beta$ for the 
disordered model (top panel) 
and the clean dimerized system (bottom panel) investigated in this study. In both panels, data sets 
are determined at the corresponding critical points and the horizontal solid lines are the theoretical 
predictions 1.09. Most of the outcomes shown in the bottom panel are 
from the results of simulations with $L=256$.}
\label{SchisT_fig1}
\end{figure}

\subsection{The universal coefficient $\Xi$}
Theoretically, a calculation with $z=1$ for the $O(N)$ nonlinear sigma model 
using the large-$N$ expansion predicts that up to the order of $1/N$, the 
quantity $\Xi$, which is defined as $S(\pi,\pi)/(\chi_s T)$, is a universal number 
given by 1.09 for $N$= 3 (which is the case here). The observables 
$S(\pi,\pi)/(\chi_s T)$ as functions of $\beta$ for the considered models 
are shown in Fig.~\ref{SchisT_fig1}. 
In both panels of Fig.~\ref{SchisT_fig1} the solid lines represent the 
theoretical value 1.09. In addition, an uncertainty of few percent (of 1.09, 
dashed lines) is included in both panels as well. The results shown in 
Fig.~\ref{SchisT_fig1} imply that although non-negligible $T$-dependence for 
$S(\pi,\pi)/ (\chi_s T)$ does appear for these models, 
the Monte Carlo data agree very well with the associated theoretical predictions. 

Most of the data shown in the bottom panel of Fig.~\ref{SchisT_fig1}, which
are associated with the clean plaquette model, are determined from
the results obtained on $L=256$ lattices. For this model, 
we have performed simulations with $L=256$ and $L=512$ for the largest value 
of $\beta$ considered ($\beta = 20$). The agreement between the results of
$S(\pi,\pi)/(\chi_s T)$ obtained from these two calculations is remarkably good
(The difference is only around one permile). Therefore the conclusion
that our Monte Carlo data are consistent with the theoretical prediction
is unquestionable

Figure \ref{SchisT_fig1} also indicates that the data of 
$S(\pi,\pi)/(\chi_s T)$ of the considered two models approach 1.0 
(dashed-dotted lines in both panels) at the regions of high temperature. 
This is consistent with the associated analytic calculations.

\begin{figure}
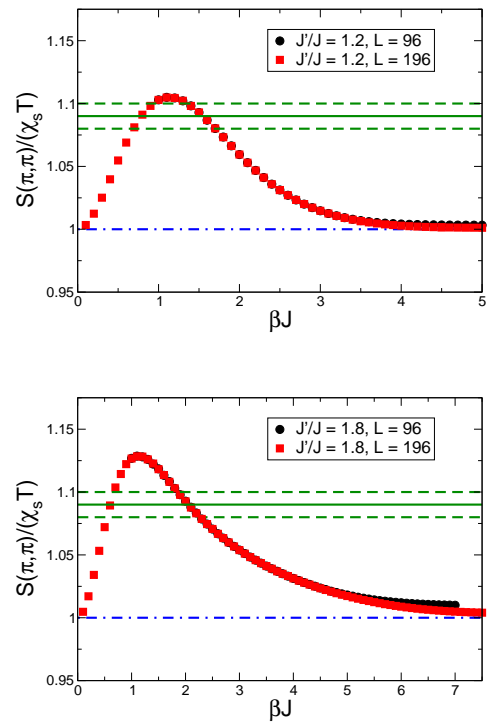

\begin{center}
\vbox{
\includegraphics[width=0.35\textwidth]{J1.2_S_chis_T_beta.eps}
\vskip0.8cm
\includegraphics[width=0.35\textwidth]{J1.8_S_chis_T_beta.eps}
}\vskip0.25cm
\end{center}
\caption{$S(\pi,\pi)/(\chi_s T)$ as functions of $\beta$ for the studied disordered model 
with $J'/J = 1.2$ (top panel) and $J'/J = 1.8$ (bottom panel). 
In both panels, the horizontal solid lines are the theoretical 
prediction 1.09. These results are calculated using 8000 (4000) MC sweeps for the 
thermalization (measurement).}
\label{SchisT_fig2}
\end{figure}

For the disordered model, in addition to the simulations performed close to the critical point, 
we have carried out calculations with $J'/J = 1.2$ and $1.8$.
The results of $S(\pi,\pi)/(\chi_s T)$ for $J'/J = 1.2$ and $1.8$ are 
demonstrated in Fig.~\ref{SchisT_fig2}. As shown in the figure, no plateaus 
appear for these two newly obtained data sets of $S(\pi,\pi)/(\chi_s T)$. 
This implies that the expected QCR behavior of this quantity does not show 
up when the calculations are conducted away from the associated QCP. 
This observed phenomenon is in agreement with the outcome determined in 
\cite{Tro98}. It is also interesting to find that at both the regions of high 
and low temperatures, the corresponding results of $\Xi$ approach 1.0 
(dashed-dotted lines in both panels of Fig.~\ref{SchisT_fig2}). This is 
again consistent with the expected theoretical prediction. 

\subsection{The universal coefficient $X$}
\begin{figure}
\vskip1.0cm
\begin{center}
\vbox{
\includegraphics[width=0.35\textwidth]{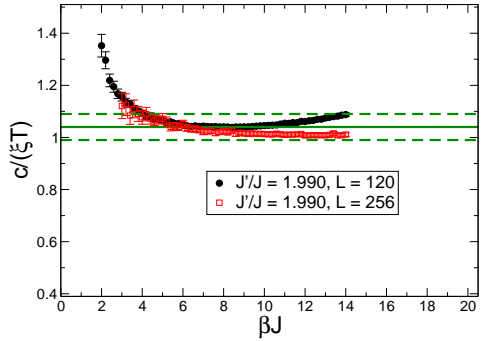}
\vskip0.7cm
}\vskip0.25cm
\end{center}\vskip-0.5cm
\caption{$c/(T\xi)$ as functions of $\beta$ for the disordered model
investigated in this study. The data sets are determined at the corresponding 
critical point $(J'/J)_c = 1.990$ and the horizontal solid line is the 
theoretical prediction 1.04.}
\label{cTxi_fig1}
\end{figure}

The final universal coefficient studied here is associated with $c/(T\xi)$ 
and is predicted to be 1.04 in theory. For the investigated disordered 
system, the associated $L=120$ and $L=256$ data of $c/(T\xi)$ as functions 
of $\beta$ are presented in Fig.~\ref{cTxi_fig1}. In the figure besides the data of 
$c/(T\xi)$, the related theoretical value and few percent error for it are 
also shown as the solid line and dashed lines, respectively. Similar to the 
scenario found in our analysis of $S(\pi,\pi)/(\chi_s T)$, a noticeable 
dependence on $T$ for the quantity $c/(\xi T)$ is observed. In addition, while 
the bulk results of the universal coefficient $X$ are reached 
only for those with $\beta < 7.5$, it is likely that for $\beta$ $\in$ $[7.5,9.0)$
the associated $X$ are the bulk ones as well. Considering the fact that 
there is a broad range of $\beta$ where the determined $X$ are within the 
theoretical predicted value with a reasonable estimated error for it, the claim
that our results shown in Fig.~\ref{cTxi_fig1} are consistent with
the outcomes conducted in Refs.~\cite{Chu93,Chu931,Chu94} is unquestionable.  
While not shown here, a similar situation occurs when $c/(\xi T)$ of the 
clean dimerized plaquette model is considered. In particular, analogous 
finite-size and finite-temperature effects as those appeared in 
Fig.~\ref{cTxi_fig1} are found.

~
~
~

\section{Discussions and Conclusions}
Using the first principles nonperturbative QMC simulations, we have investigated the 
exotic characteristics of QCR related to both a 2D spin system with 
configurational disorder and a 2D clean dimerized spin-$\frac{1}{2}$ Heisenberg 
model. These unique properties of the considered models result from the 
interplay of the thermal and the quantum fluctuations. 
We firstly reconfirm that the dynamic exponent $z$ for the disordered
model studied here is 1. With this result, as well as the fact that $z=1$ for
the clean dimerized plaquette model, the three universal coefficients associated 
with QCR, namely $\Omega$, $\Xi$, and $X$ are calculated. 
We find our Monte Carlo data of both the disordered and the clean systems are 
consistent with the analytic results based on the large-$N$ calculations of the 
$O(N)$ nonlinear sigma model. It is interesting to notice that while quantum 
systems with certain kinds of quenched disorder, such as the configurational 
disorder employed in this study, violate the Harris criterion 
\cite{Har74,Cha86,Mot00,Yao10,Nvs14}, 2D disordered spin-$\frac{1}{2}$ models 
with bond dilution fulfill this principle \cite{Vaj02,San02,Skn04,Yu05,San06}. 
The results presented here seem to imply the scenario that disordered 
systems which violate the Harris criterion conform the theoretical predictions 
of QCR. It will be compelling to investigate whether for models 
satisfy the Harris criterion, the corresponding values of 
the three universal coefficients of QCR remain the same as the known ones
in the literature.   

While the numerical data obtained from the QMC simulations are in good 
agreement with the corresponding analytic predictions, non-negligible 
dependence on $T$ is observed for these three universal coefficients. 
Furthermore, for both the considered models, the estimated values of $\Omega$, 
which is related to $\chi_u c^2$, are different statistically from the 
analytic and numerical ones established in the literature (except that
determined in Ref.~\cite{Sen15}). 
The difference between the values of $\Omega$ 
estimated here ($\Omega = 0.306(10)$) and the theoretical result previously known 
($\Omega \sim 0.27$) is more than 10 percent, which cannot be accounted for by 
the potential systematic uncertainties resulting from the calculations of 
$c$ conducted in this study. Among the relevant studies associated with $\Omega$, 
only the dedicated work of Ref.~\cite{Sen15} agrees with ours. 
It is also interesting to notice that the $\Omega$ 
estimated in Ref.~\cite{Sen15} is somewhat (slightly) larger than what has been 
calculated here. We attribute this
to the fact that data with temperatures (lattice sizes) lower (larger) than ours 
were used in that work for the related analysis. 
Indeed, in our investigation with a fixed $z=1$, the magnitude of $\Omega$
is increasing when more and more data of higher $T$ are excluded in the fits.
Aside from that, finite-temperature effect clearly shows up for $\chi_uc^2/T$, as can 
been seen in Figs.~\ref{omega_fig4} and \ref{omega_fig7}. Such an effect to 
some extent will influence the determination of $\Omega$ if the formula 
$\chi_u c^2 = \Omega T$ is used to extract the value of $\Omega$.

It is intriguing that while the analytic outcome of $\Omega$ including both
the leading and subleading contributions deviates significantly from its 
numerical estimations obtained in this study and in Ref.~\cite{Sen15},
the theoretical prediction of $\Omega$ by considering only the leading term
is in better agreement with our results and that of Ref.~\cite{Sen15}. 
In summary, the numerical evidence reached 
here for the described discrepancy is quite convincing. A detailed
study of $\chi_u$ for the clean bilayer spin-$\frac{1}{2}$ model demonstrates
such a deviation as well \cite{Sen15}. To shed light on 
this deviation, besides conducting analytic studies associated with 
corrections not considered before, it will be desirable as well to simulate 
other disordered and clean dimerized models other than those investigated here.


\section{Acknowledgments}
\vskip-0.3cm
We thank A.~W.~Sandvik for bringing Ref.~\cite{Sen15}, in which
a detailed investigation of $\Omega$ based on a clean bilayer
spin model was conducted, to our attention. 
This study is partially supported by Ministry of Science and 
Technology of Taiwan.


\begin{thebibliography}{99}

\bibitem{Cha88}
S.~Chakravarty, B.~I.~Halperin, and D.~R. Nelson,
Phys. Rev. Lett. {\bf 60}, 1057 (1988).


\bibitem{Reg88}
J.~D.~Reger and A.~P.~Young, Phys. Rev. B {\bf 37}, 5493 (1988).

\bibitem{Cha89}
S.~Chakravarty, B.~I.~Halperin, and D.~R. Nelson,
Phys. Rev. B {\bf 39}, 2344 (1989).


\bibitem{Oit941}
J.~Oitmaa, C.~J.~Hamer, and Zheng Weihong, Phys. Rev. B {\bf 50}, 3877 (1994).

\bibitem{Oit942}
C.~J.~Hamer, Zheng Weihong, and J.~Oitmaa, Phys. Rev. B {\bf 50}, 6877 (1994).

\bibitem{Wie94}
U.-J.\ Wiese and H.-P.\ Ying, Z.\ Phys.\ B {\bf 93}, 147 (1994).

\bibitem{San951}
A.~W.~Sandvik and D.~J.~Scalapino, Phys. Rev. B {\bf 51}, 9403 (1995).\

\bibitem{Bea96}
B.\ B.\ Beard and U.-J.\ Wiese, Phys.\ Rev.\ Lett.\ {\bf 77} (1996) 5130.

\bibitem{San97}
A.\ W.\ Sandvik, Phys. Rev. B {\bf 56}, 18 (1997).

\bibitem{Jia08}
F.-J.~Jiang, F.~Kampfer, M.~Nyfeler, and U.-J. Wiese, Phys. Rev.
B {\bf 78}, 214406 (2008).

\bibitem{Jia11.1}
F.-J.~Jiang and U.-J.Wiese, Phys. Rev. B {\bf 83}, 155120 (2011).


\bibitem{Chu93}
A.~V.~Chubukov and S.~Sachdev, Phys. Rev. Lett. {\bf 71}, 169 (1993).

\bibitem{Chu931}
A. V. Chubukov and S. Sachdev, Phys. Rev. Lett. {\bf 71}, 2680 (1993).

\bibitem{Sok94}
Alexander~Sokol, Rodney~L.~Glenister, and Rajiv~R.~Singh,
Phys. Rev. Lett. {\bf 72}, 1549 (1994).


\bibitem{Chu94}
A.~V.~Chubukov, S.~Sachdev, and J.~Ye, Phys. Rev. B {\bf 49}, 11919 (1994).

\bibitem{San95}
A.~W.~Sandvik, A.~V.~Chubukov, and S.~Sachdev, Phys. Rev. B {\bf 51}, 16483 (1995)

\bibitem{Tro96}
M.~Troyer, H.~Kantani, and K.~Ueda, Phys.~Rev.~Lett. {\bf 76}, 3822 (1996).

\bibitem{Tro97}
Matthias Troyer, Masatoshi~Imada, and Kazuo~Ueda, J. Phys. Soc. Jpn. 66, 2957 (1997).

\bibitem{Tro98}
Jae-Kwon Kim and Matthias~Troyer, Phys.~Rev.~Lett. {\bf 80}, 2705 (1998).

\bibitem{Kim99}
Y.~J.~Kim, R.~J.~Birgeneau, M.~A.~Kastner, Y.~S.~Lee, Y.~Endoh, G.~Shirane, and K.~Yamada,
Phys. Rev. B 60, 3294 (1999).

\bibitem{Kim00}
Y.~J.~Kim and R.~J.~Birgeneau, Phys. Rev. B {\bf 62}, 6378 (2000).

\bibitem{Yao10}
Dao-Xin Yao, Jonas Gustafsson, E.~W.~Carlson, and Anders W.~Sandvik,
Physical Review B, {\bf 82}, 172409 (2010).

\bibitem{Sen15}
A. Sen, H. Suwa, and A. W. Sandvik, Phys. Rev. B {\bf 92}, 195145 (2015).

\bibitem{San99}
A.~W.~Sandvik, Phys. Rev. B {\bf 66}, R14157 (1999).

\bibitem{San02}
A.~W.~Sandvik, Phys. Rev. B {\bf 66}, 024418 (2002).

\bibitem{rhosL_def}
The quantity $\rho_s$ (spin stiffness) is defined as 
$\rho_s = \frac{1}{2\beta}\sum_{i=1,2}\langle W_i^2\rangle$,
where $W_i$ is the winding number in the spatial $i$-direction.

\bibitem{Wen09} 
S. Wenzel and W. Janke, Phys. Rev. B {\bf 79}, 014410 (2009).


\bibitem{Jia11}
F.-J. Jiang, Phys. Rev. B {\bf 83}, 024419 (2011).

\bibitem{size_convergence}
The $\chi_u c^2/T$ data of the clean plaquette model determined at $(J'/J)_c$ are examined 
carefully and indeed those shown in this study are size convergent.
It is anticipated that the results associated with $J'/J = 1.8228$ and 1.8232
should be size convergent as well.


\bibitem{Nvs14}
Nvsen Ma, Anders W. Sandvik, and Dao-Xin Yao, Phys. Rev. B {\bf 90}, 104425 (2014).

\bibitem{Har74}
A.~B.~Harris, J. Phys. C {\bf 7}, 1671 (1974).

\bibitem{Cha86}
J.~T.~Chayes, L.~Chayes, D.~S.~Fisher, and T.~Spencer, Phys. Rev.
Lett. {\bf 57}, 2999 (1986).

\bibitem{Mot00}
O. Motrunich, S.C. Mau, D.A. Huse, and D.S. Fisher, 
Phys. Rev. B {\bf 61}, 1160 (2000).

\bibitem{Vaj02}
O.~P.~Vajk and M.~Greven, Phys. Rev. Lett. {\bf 89}, 177202 (2002).

\bibitem{Skn04}
R. Sknepnek, T. Vojta, and M. Vojta, Phys. Rev. Lett. {\bf 93},
097201 (2004).

\bibitem{Yu05}
Rong Yu, Tommaso Roscilde, and Stephan Haas, Phys. Rev. Lett. {\bf 94}, 
197204 (2005)

\bibitem{San06}
A.~W.~Sandvik, Phys. Rev. Lett. {\bf 96}, 207201 (2006).


\end{thebibliography}
\end{document}